\documentclass[twocolumn]{aastex631}

\usepackage{mathtools}
\let\splitbox\undefined

\graphicspath{{./}{figures/}}
\usepackage{graphicx}
\usepackage{subcaption}
\usepackage[utf8]{inputenc}
\usepackage[export]{adjustbox}

\usepackage{natbib}
\usepackage{color}

\usepackage{wrapfig}

\definecolor{darkorange}{rgb}{0.7, 0.3, 0.}

\begin{document}

\title{3D solar coronal loop reconstructions with machine learning}

\author[0000-0002-5481-9228]{Iulia Chifu}
  \affiliation{Max-Planck-Institut f\"ur Sonnensystemforschung, \\
        Justus-von-Liebig-Weg 3, 37077 G\"ottingen, Germany}
      
\affiliation{Astronomical Institute of Romanian Academy, \\
    Cutitul de Argint 5, Bucharest, Romania}

\author[0000-0003-4920-0153]{Ricardo Gafeira}
\affiliation{Univ Coimbra, IA, CITEUC, OGAUC, Coimbra, Portugal\\
            Rua do Observatório s/n,3040-004 Coimbra, Portugal}
\affiliation{         Instituto de Astrof\'{i}sica de Andaluc\'{i}a (CSIC),\\ Apartado de Correos 3004, E-18080 Granada, Spain}
\email{chifu@mps.mpg.de}

\begin{abstract}
The magnetic field plays an essential role in the initiation and evolution of different solar phenomena in the corona. The structure and evolution of the 3D coronal magnetic field are still not very well known. A way to get the 3D structure of the coronal magnetic field is by performing magnetic field extrapolations from the photosphere to the corona. In previous work, it was shown that by prescribing the 3D reconstructed loops' geometry, the magnetic field extrapolation finds a solution with a better agreement between the modeled field and the  reconstructed loops. Also, it improves the quality of the field extrapolation. Stereoscopy represents the classical method for performing 3D coronal loop reconstruction. It uses at least two view directions. When only one vantage point of the coronal loops is available, other 3D reconstruction methods must be applied. Within this work, we present a method for the 3D loop reconstruction based on machine learning. Our purpose for developing this method is to use as many observed coronal loops in space and time for the modeling of the coronal magnetic field. Our results show that we can build machine learning models that can retrieve 3D loops based only on their projection information. In the end, the neural network model will be able to use only 2D information of the coronal loops, identified, traced and extracted from the EUV images, for the calculation of their 3D geometry.

\end{abstract}

\keywords{3D reconstruction, coronal loops}


\section{Introduction} \label{sec:intro}

Coronal loops are approximated to a 3D arch-like structure that has their footpoints located in the opposite polarity of the magnetic field. In the solar corona, the plasma pressure is dominated by the magnetic pressure \citep{Gary2001} which makes the coronal loop plasma confined in the 3D flux tubes. The coronal loops are mainly visible on the solar disk in the EUV images which shows only the projection of the 3D structures. The reconstruction of the three-dimensional shape of coronal loops is important for the implication on the magnetic field and topology \citep{2010LRSP....7....5R}.

The magnetic field plays an important role in the initiation and evolution of different solar phenomena in the corona. While at the photospheric level full-disk vector magnetic field measurements are provided by several observatories, in the chromosphere and corona the vector magnetic field is only sometimes provided and only for small fields of views. Different approaches are used to obtain the coronal magnetic field. One of the approaches is based on magnetic field extrapolations into the corona having as bottom boundary photospheric magnetic field data (see \cite{WiegelmannSakurai2012, 2015SSRv..tmp...75W}, for reviews). Another approach is a stereoscopic reconstruction based on simultaneously recorded images from at least two view directions in the EUV wavelengths (chromosphere and low corona) and white light (high corona) (see \cite{2011LRSP....8....5A} for a review). When applied for the 3D reconstruction of the coronal magnetic loops, the output of the two methods does not match \citep{2009ApJ...696.1780D}. \cite{2017ApJ...837...10C} showed that by constraining the nonlinear force-free field (NLFFF) extrapolations with the 3D stereoscopically reconstructed loops, one can obtain a better field model. 

The 3D stereoscopic reconstruction method requires images of the same object recorded from at least two view directions. This requirement limits the time range for finding the 3D information on solar coronal loops. One of the steps for stereoscopy is identification and matching of the same object in the available images \citep{Inhester2006}. This condition limits the possibility of selecting large number of coronal loops.  

With the method laid out in this work, we can make use of all of the epochs at which exists simultaneously EUV images and vector magnetograms. The 2D loops which one would be able to identify, visually or automatically, in an EUV image, will be 3D reconstructed by the neural network method. The resulted 3D loops can be therefore used as a constrain for the magnetic field extrapolation as shown by \cite{2017ApJ...837...10C}. This method allows us to use many more loops as a constraint for extrapolation than was possible till now. The method we present here for the 3D coronal loop reconstruction will be able to constrain the coronal magnetic field models to fit the observations. These coronal magnetic field models can be further used by the MHD models in the attempt to investigate the heating mechanisms of the coronal loops. 

\section{Methods}
\subsection{Convolution Neural Network}
\label{CNN_description}

Neural networks (NN) are designed to work like the processes inside a human brain. In this context, what we call a neuron is a mathematical function that infers quantities based on the input data according to the NN architecture.

Among the various different NN architectures, there is a subset called convolutional neural network (CNN). These networks use convolutional layers where one or two-dimensional filters are applied to the input data. The unidimensional (1D) CNN basic structure used in this work can be then resumed as a combination of the following layers: 1.) layers that apply a convolution between a kernel and each set of data points (in our case loop position), with the goal of enhancing particular features in the data; 2.) pooling layer that downsamples the data by reducing the data information but keeping the more prominent features; 3.) Finally, the so-called fully connected layers which are taking the results of the previous operations and multiply them by a set of weights to estimate the final output. Most of the operations performed in these layers have an associated activation function, which are mathematical equations that determines the importance of a certain variation.

One key element of any NN is the training process in which an input data set and its respective output data are used. This allows the model to determine the weights that the fully connected layers will apply. The network uses an optimization procedure to minimize the differences, using a loss function, between the training set and the estimated output. It is common practice to use a fraction of the input data not used for the training and compare it to the model output. That is called validation loss. It allows us to evaluate the evolution of the training and identify problems like overfitting. To optimize the minimization process, we can divide the data into small batches. During the training, the weights will be updated at the end of each batch, leading to a better and faster convergence.
The overfitting can be avoided by introducing dropout elements that randomly remove a fraction of the points from the training process introducing a level of randomness to the system and helping it to avoid false local minima. We call epoch each full iteration of the training process described before.

In this work, we follow the CNN architecture model presented on \cite{gafeira2021}. We optimized the hyperparameters by trial-and-error until we found a CNN model satisfying a loss function being at least 10$^{-3}$ of the normalization values to all considered cases.
We trained the model using the optimizer AdaMax witch is a first-order gradient-based optimization of stochastic objective functions \citep{2014arXiv1412.6980K}. We used a batch size of 64 and a validation split of 0.3. The convolutional layers and the first dense layer use the nonlinear function called the rectified linear activation unit (ReLU) \citep{Nair:2010:RLU:3104322.3104425} and the last dense layer a linear function. The global structure of the CNN and the respective hyperparameters used in this work is summarised in Table \ref{cnn}.

The input parameters are the 2D position of the loops (X, Y), the length of the loop, the distance between the footpoints of the loops, and the angle defined by the two footpoints with the top of the loops. The 2D positions X and Y are interpolated to all have a fixed size of 1500 points.


The output and training set of the CNN is the Z component of the corresponding 2D loop with a fixed size of 1500 points. 

We applied a normalization factor that corresponds to the loop average height used in the training of CNN. In this work the normalization used is 8.75 ADU for Data type 1, and for Data type 2 Case 1 is 10.1 Mm, Case 2 is 77.54 Mm and Case 3 is 73.15 Mm.

\begin{table}[h!]
\centering
 \begin{tabular}{||c|c|c||} 
 \hline
  Layer & Size&Activation\\
 \hline
  \hline
  1D convolutional &  150&ReLU\\
 \hline
  1D MaxPooling &  2&\\
 \hline
  1D convolutional & 90 & ReLU\\
 \hline
   1D MaxPooling & 2 &  \\
 \hline
   1D convolutional & 45 & ReLU\\
 \hline
   Dropout & factor of 0.25 &\\
 \hline
   Fully connected  &  4 times the loop &\\ 
   dense layer &length & ReLU\\
 \hline
    Fully connected & loop length &\\ dense layer& length & linear \\
 \hline
\end{tabular}
\caption{CNN model architecture used in this work.}
\label{cnn}
\end{table}

\subsection{Nonlinear Force-Free Field extrapolation}
To produce a big enough data set for training the CNN we used the nonlinear force-free field extrapolation method, proposed initially by \cite{WheatlandEtal2000} and implemented by \cite{Wiegelmann04}. The NLFFF optimization method is based on the minimization of a functional L$_\text{tot}$  as a sum of the following terms 
\begin{eqnarray}
  \text{L}_1=\int_V w_f \frac{|(\nabla \times \mathbf{B}) \times
    \mathbf{B}|^2}{B^2} \;d^3r,
  \label{L1}\\
  \text{L}_2=\int_V w_f |\nabla \cdot \mathbf{B}|^2 \;dr^3,
  \label{L2}\\
  \text{L}_3=\int_S ( \mathbf{B} - \mathbf{B}_{obs} )
           \cdot \mathrm{diag(\sigma^{-2}_\alpha)}
           \cdot (\mathbf{B}-\mathbf{B}_{obs}) \;d^2r,
  \label{L3}
\end{eqnarray}
where the minimization of $L_1$ satisfies the force-free condition and the $L_2$ satisfies the Gauss Law. The minimization of the $L_3$ ensures that at the photospheric level the modeled field ($\mathbf{B}$) is as close as possible to the observed one ($\mathbf{B_{obs}}$). $w_f$  is the boundary weight function \citep[more details in][]{Wiegelmann04}, $\mathrm{diag(\sigma^{-2}_\alpha)}$ represents the estimated measurement errors for the three field component \citep[see][for more details]{TadesseEtal2011}. The method was implemented in the Cartesian \citep{Wiegelmann04} and spherical \citep{TadesseEtal2011} coordinate system.  Both of the implementations are using as a bottom boundary the vector magnetic field. The lateral and top boundaries are fixed from an initial calculated potential field.

\section{Data preparation}
We tested the CNN method in four different cases by using input loops obtained from different solutions of the NLFFF extrapolation. The first case uses synthetic data and the last three are based on observational data:
\begin{enumerate}
    \item \textbf{Data type 1}. The NLFFF extrapolation uses as bottom boundary a semi-analytical force-free field solution proposed by \cite{1990ApJ...352..343L}. The solution is obtained by solving the Grad-Shafranov equation for an axisymmetric nonlinear force-free field. The results obtained based on the synthetic data are defined as Case 0.
    \item \textbf{Data type 2}. In this category are the data obtained by the Heliospheric Magnetic Imager (HMI) instrument \citep{2012SoPh..275..285C} onboard Solar Dynamic Observatory (SDO) spacecraft \citep{2012SoPh..275....3P}. We defined three Cases:
        \begin{itemize}
            \item Case 1. We use as input for the NLFFF a field of view of 1280$\times$400 from a full disk HMI vector magnetogram.
            \item Case 2. For the NLFFF extrapolation method in spherical coordinates, we used as bottom boundary a full disk vector magnetogram recorded on 02 August 2010.
            \item Case 3. For the last case, we use synoptic vector magnetograms which are HMI data products suitable for the full Sun extrapolation \citep{Liu2017}. These maps are constructed from the daily HMI vector magnetograms. The size of the HMI synoptic magnetograms is 3600$\times$1440 pixels and the pixel size is 0.1 deg in longitude and 0.001 in sin latitude. 
        \end{itemize}
 \end{enumerate}

From the solutions of the extrapolations, using a fourth-order Runge-Kutta method, we traced 3D magnetic field lines and we projected them on the solar surface. 

\section{Results}
\label{sec:results}
\subsection{Data type 1}
\label{DataType1}
\textit{Case 0}. We tested the CNN method described in Section \ref{CNN_description} with 3D and 2D loops obtained from an NLFFF extrapolation which used as bottom boundary a Low \& Lou semi-analytical force-free field solution. We used the Cartesian coordinate system for a computational box in the X, Y, and Z-direction of 192$\times$192$\times$96 grid points. For the training of the CNN, we used 21704 loops traced from the NLFFF solution.

\begin{figure*}[ht]
\begin{subfigure}{0.5\textwidth}
\includegraphics[width=0.3\textheight, trim = 0 0 0 10, clip]{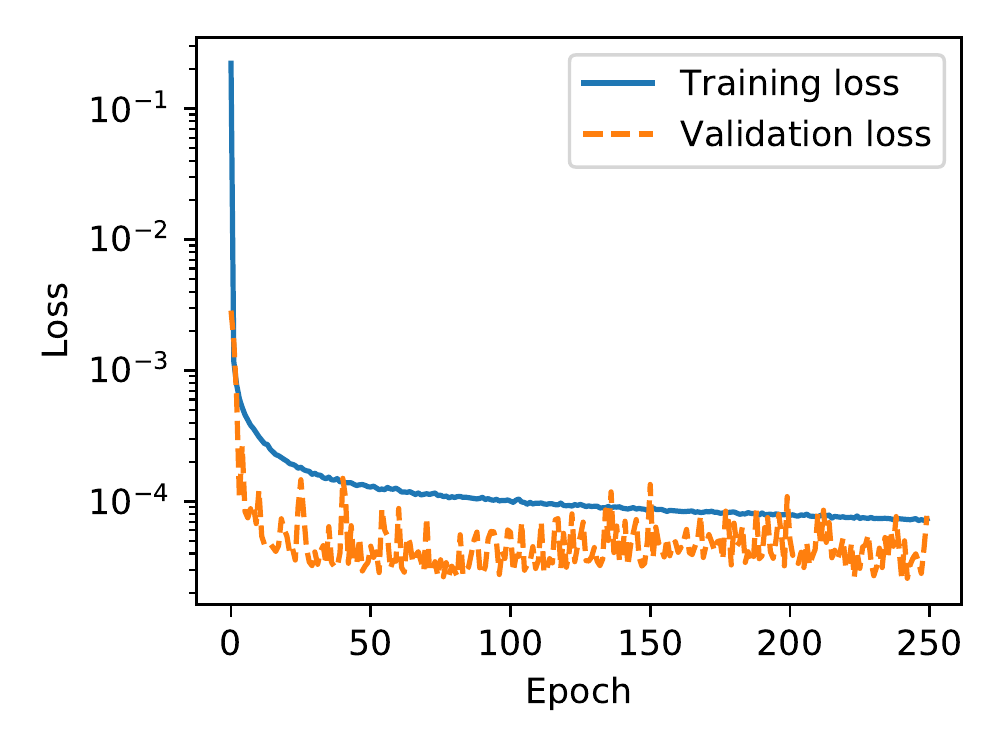}
\caption{Display of the training (blue solid line) and validation (orange dashed line) loss function during the CNN training.}
\label{fig:Case0a}
\end{subfigure}
\begin{subfigure}{0.5\textwidth}
\includegraphics[width=0.3\textheight, trim = 40 240 20 250, clip]{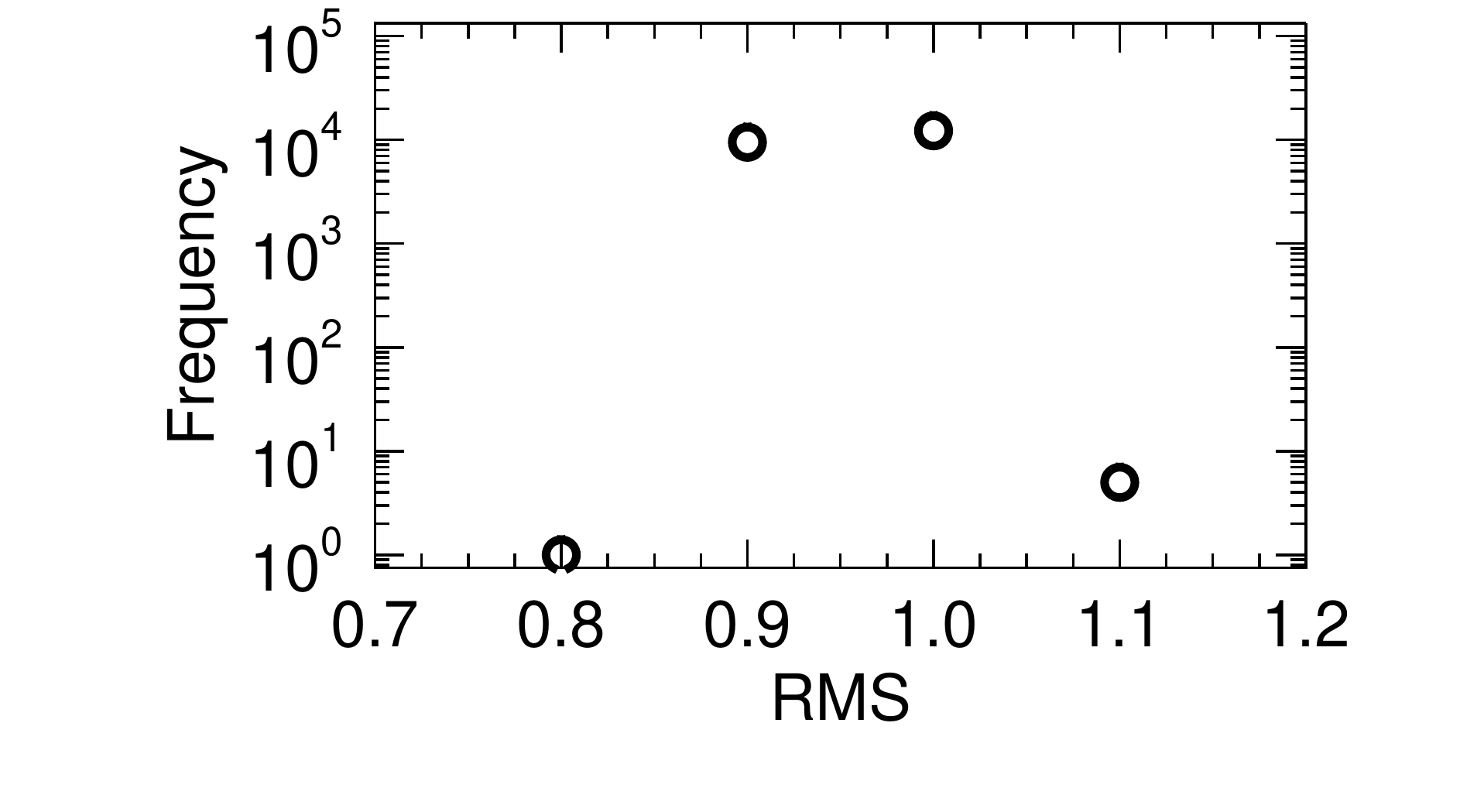}
\caption{Distribution of the root mean square for Case 0 loops.}
\label{fig:Case0b}
\end{subfigure}
\begin{subfigure}{\textwidth}
\hspace{3cm} \includegraphics[width=0.45\textheight, trim = 20 10 40 0, clip]{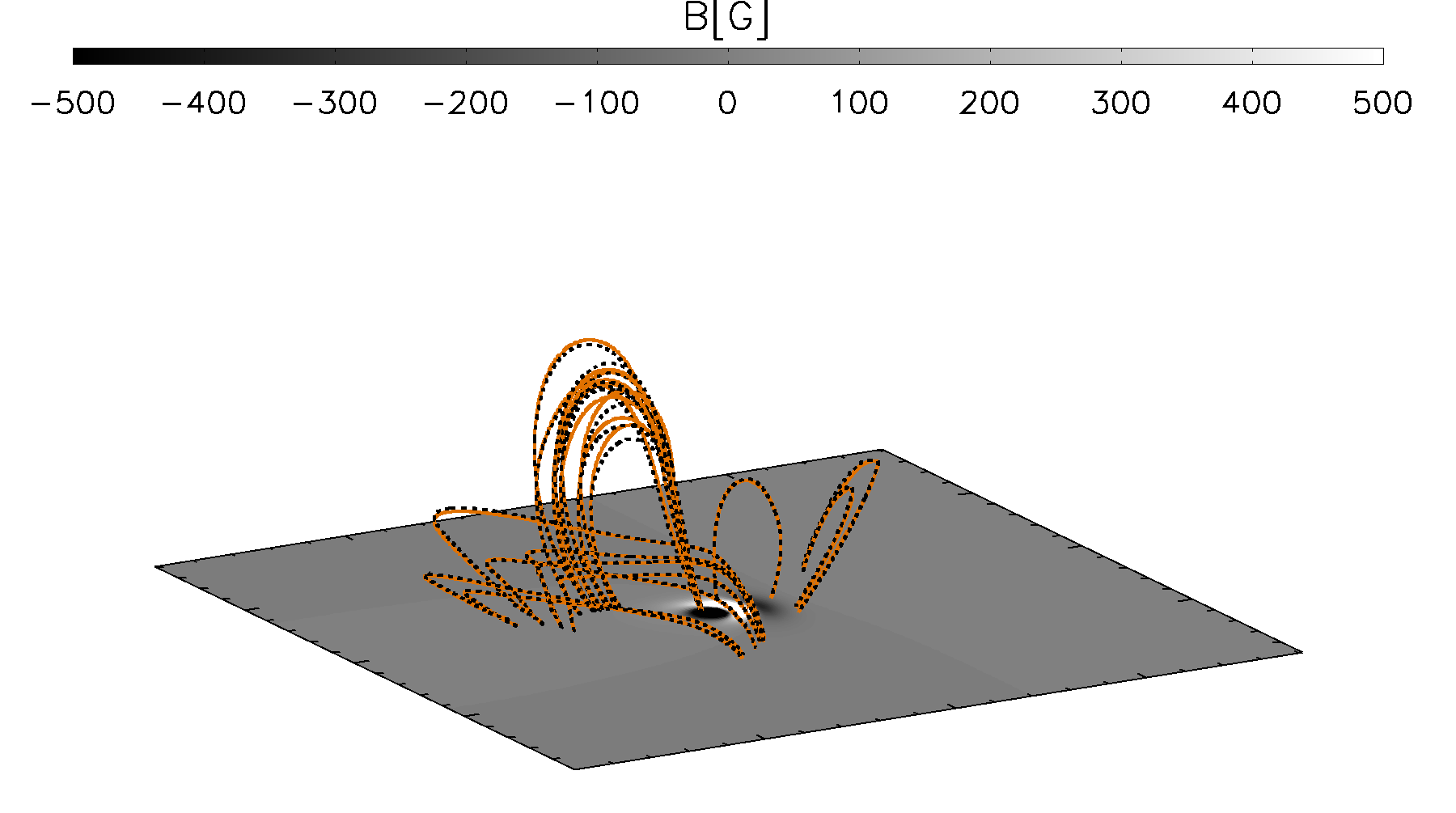}
\caption{3D CNN reconstructed loops with solid orange lines and original 3D loops with black dashed lines overplotted on the Low \& Lou bottom boundary.}
\label{fig:Case0c}
\end{subfigure}

\caption{CNN results for Data type 1, Case 0.}
\label{fig:image0}
\end{figure*}

We trained the CNN initially using 250 epochs. After the training, we evaluate the result of the training and validation loss function. 
Since the model has converged after 250 epochs, we stopped the training process. In the presented model we did not find any indication of overfitting. That would be visible as a sudden increase of the validation loss function. In Fig.\ref{fig:Case0a} we display the validation and training loss evolution obtained from the CNN model for Case 0.

For the evaluation of the CNN solution, we calculated the root mean square 
\begin{equation}
    RMS = \sqrt{\frac{\sum\mathfrak{R^2}}{N}},
    \label{RMS}
\end{equation}

 where $\mathfrak{R}$ is the ratio between the Z coordinate of the reconstructed and the original loops and N is the number of elements in a loop. 
 
 In Fig.\ref{fig:Case0b} we show the RMS distribution for Case 0. The RMS values spanning 0.9 to 1.1 represent 99.96\% from the entire data set of loops. An RMS value of unity means the CNN reconstructed loop and the original ones are perfectly matched. The CNN could obtain almost perfectly the 3D position for 99.96\% of loops. In Fig. \ref{fig:Case0c} we display the Low \& Lou magnetic field solution used for the NLFFF extrapolation overplotted with a randomly selected subset (for better visibility) of the original traced and the reconstructed 3D loops. 

\subsection{Data type 2}
\label{DataType2}
\begin{figure*}[ht]
\begin{subfigure}{0.5\textwidth}
\includegraphics[width=0.3\textheight, trim = 0 0 0 10, clip]{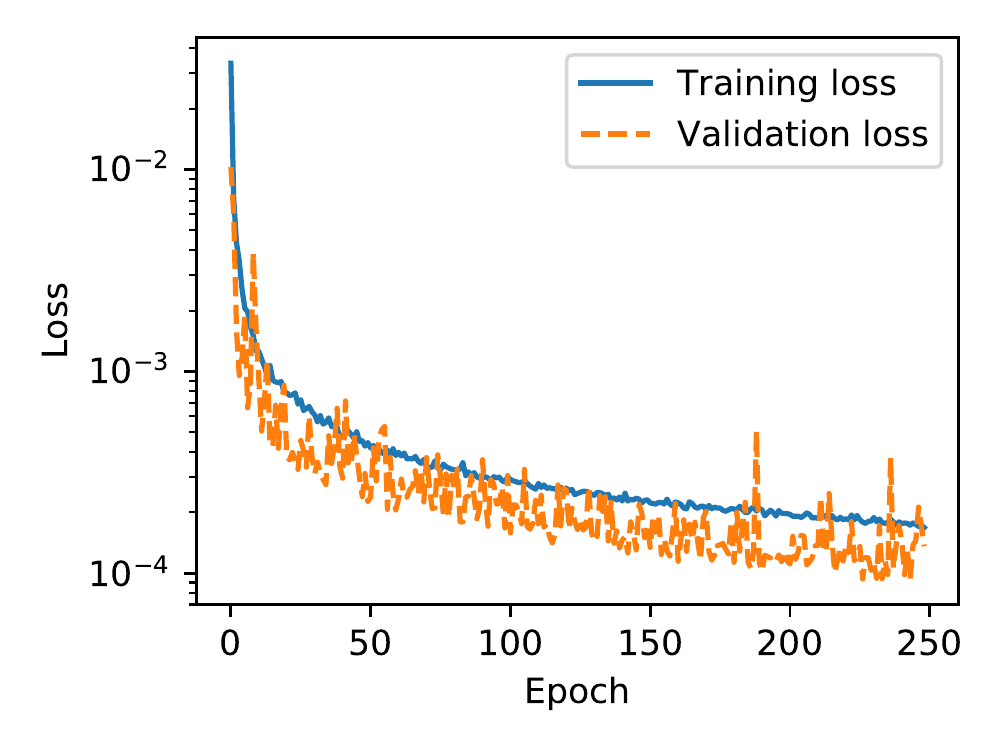}
\caption{Display of the training (blue solid line) and validation (orange dashed line) loss function during the CNN training.}
\label{fig:Case1a}
\end{subfigure}
\begin{subfigure}{0.5\textwidth}
\includegraphics[width=0.3\textheight, trim = 40 240 20 250, clip]{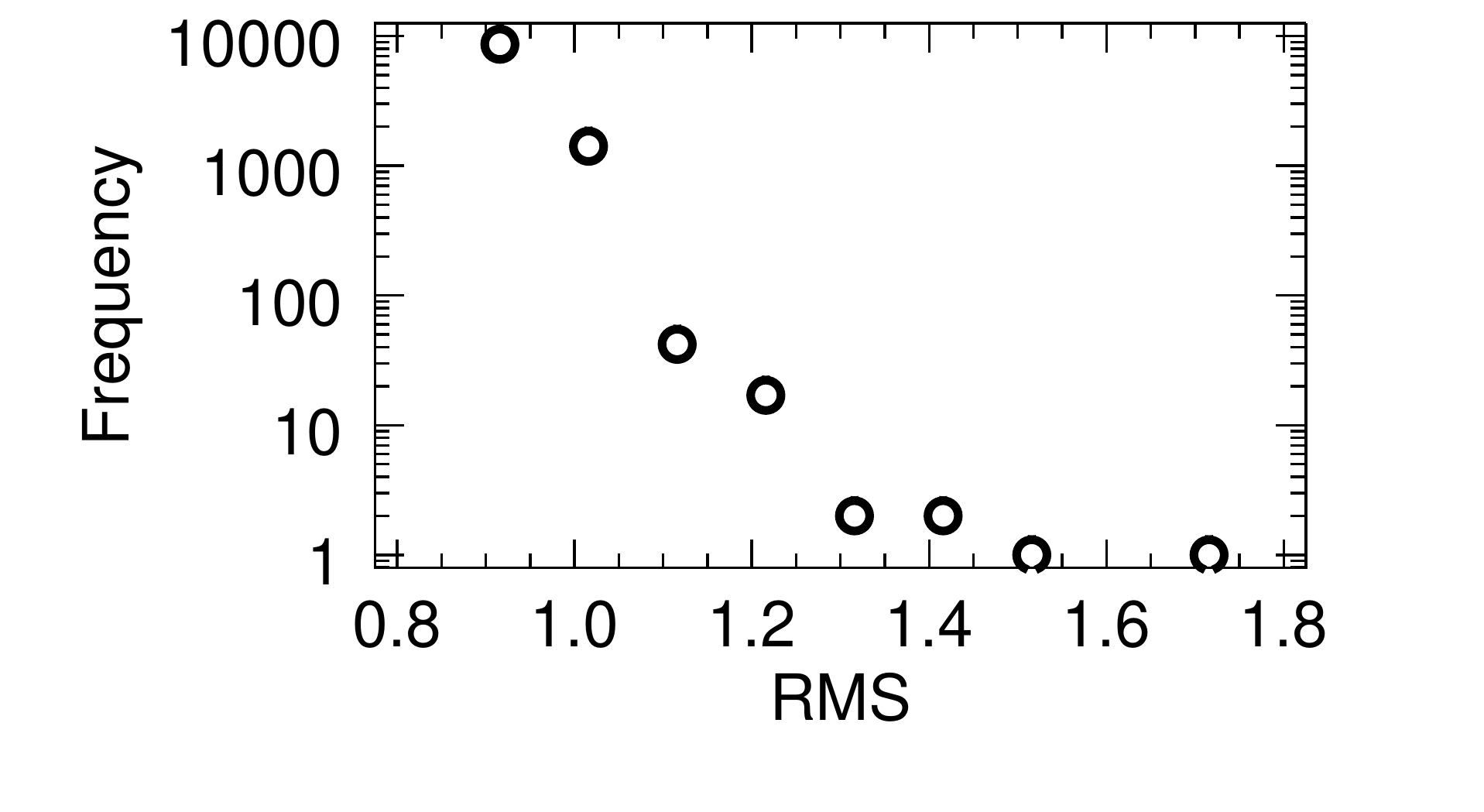}
\caption{Distribution of the root mean square for Case 1 loops.}
\label{fig:Case1b}
\end{subfigure}
\begin{subfigure}{\textwidth}
\hspace{3cm} \includegraphics[width=0.5\textheight, trim = 20 20 0 10, clip]{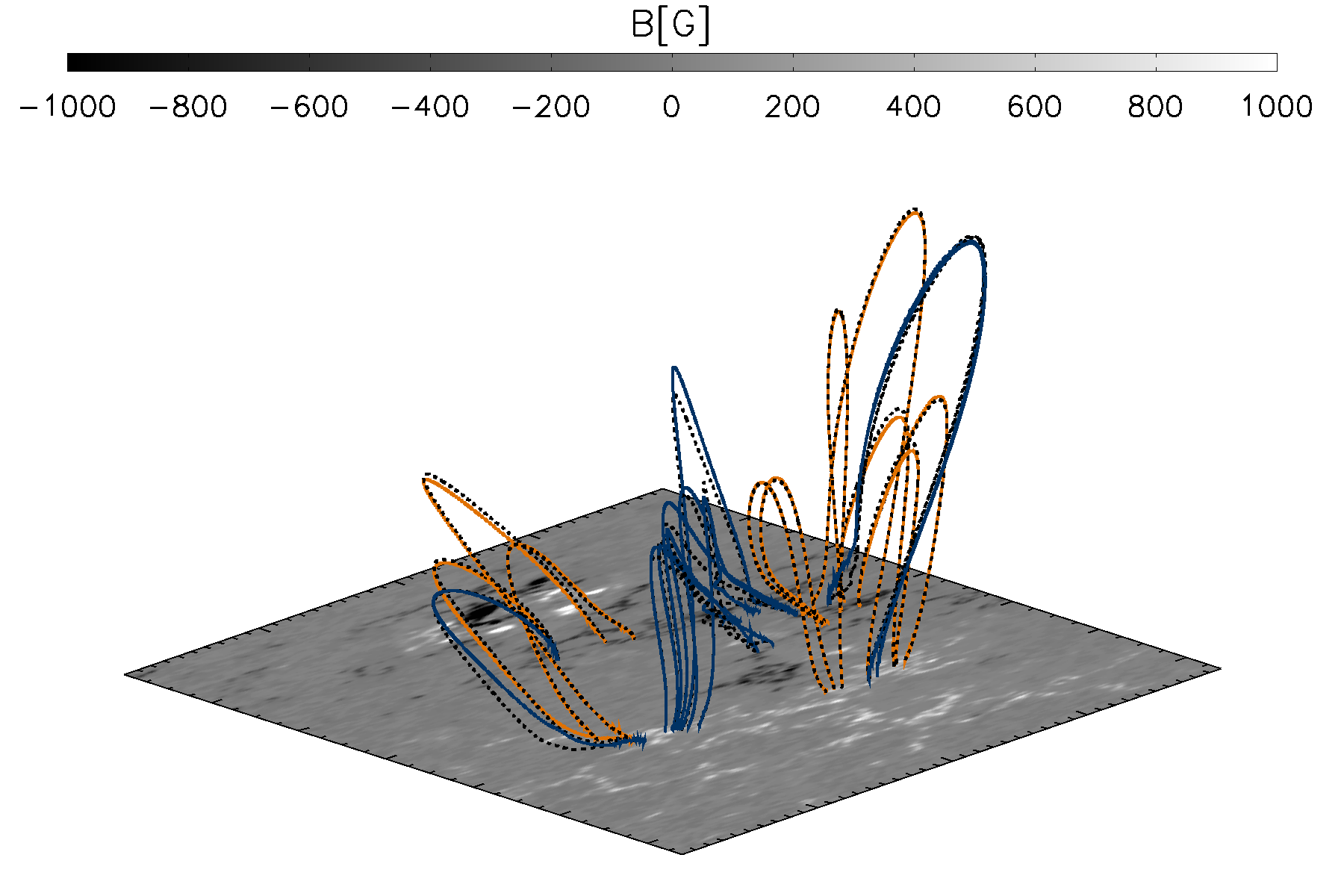}
\caption{Randomly selected 3D loops overplotted on the HMI phototspheric radial magnetic field. The black dashed lines shows the original 3d loops. The CNN reconstructed 3D loops with an RMS smaller than 1.1 are displayed with solid orange lines. In blue solid line are shown the 3D loops with an RMS larger than 1.1. }
\label{fig:Case1c}
\end{subfigure}

\caption{CNN results for Data type 2, Case 1.}
\label{fig:image1}
\end{figure*}

\textit{Case 1}. We applied the NLFFF extrapolation in the Cartesian geometry to an HMI vector magnetic field recorded on 15 July 2010. From the full disk 4096$\times$4096 pixels we selected a 1280$\times$400 field of view. The height of the box is 320 pixels. From the solution of the extrapolation, we traced 10127 loops. The CNN network reached a reasonable solution after 250 epochs similar to Case 0. In Fig. \ref{fig:Case1a} we plot the evolution of the training and validation losses with the epochs. Fig. \ref{fig:Case1b} shows the distribution of the RMS (Eq. \ref{RMS}). The RMS values between 0.9 and 1.1 represent a percentage of 99.06 of all loops. In Fig.\ref{fig:Case1c} we present the HMI vector magnetogram used as input for extrapolation together with few randomly selected 3D original and reconstructed loops (orange and blue solid line). The two different categories of reconstructed loops are made based on the RMS values. The orange loops are the ones with RMS smaller than 1.1 and the blue loops with RMS larger than 1.1.

\textit{Case 2}. For the spherical NLFFF optimization code we use full disk HMI vector magnetic field recorded on 02 August 2010. The computational domain of 256$\times$372$\times$512 grid points extends over r = [1, 2.5] R$_\odot$, $\theta$ = [-70$^\circ$, 70$^\circ$] in latitude and $\phi$ = [90$^\circ$, 270$^\circ$] in longitude. The latitudinal boundaries exclude the polar areas because the surface data in polar latitudes have poor quality and the numerical finite-difference representation used for the $L_{tot}$ functional expressed in the spherical coordinates becomes singular at the poles. From the solution of the extrapolation, we traced and used for the CNN 10368 3D loops. 98.31\% of the loops had an RMS (Eq. \ref{RMS}) between 0.5 and 1.5. 

\textit{Case 3}. We used as boundary data full sun synoptic vector magnetogram.

\begin{figure*}[ht]
\begin{subfigure}{0.5\textwidth}
\includegraphics[width=0.3\textheight, trim = 0 0 0 10, clip]{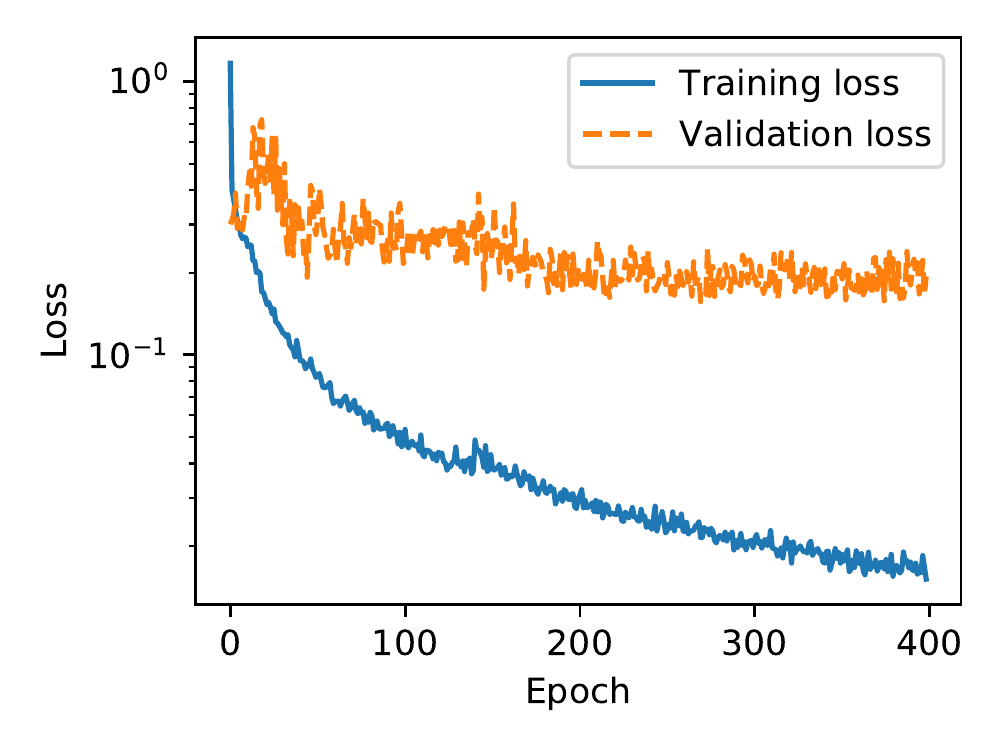}
\caption{Display of the training (blue solid line) and validation (orange dashed line) loss function during the CNN training.}
\label{fig:Case3a}
\end{subfigure}
\begin{subfigure}{0.5\textwidth}
\includegraphics[width=0.3\textheight, trim = 40 240 20 250, clip]{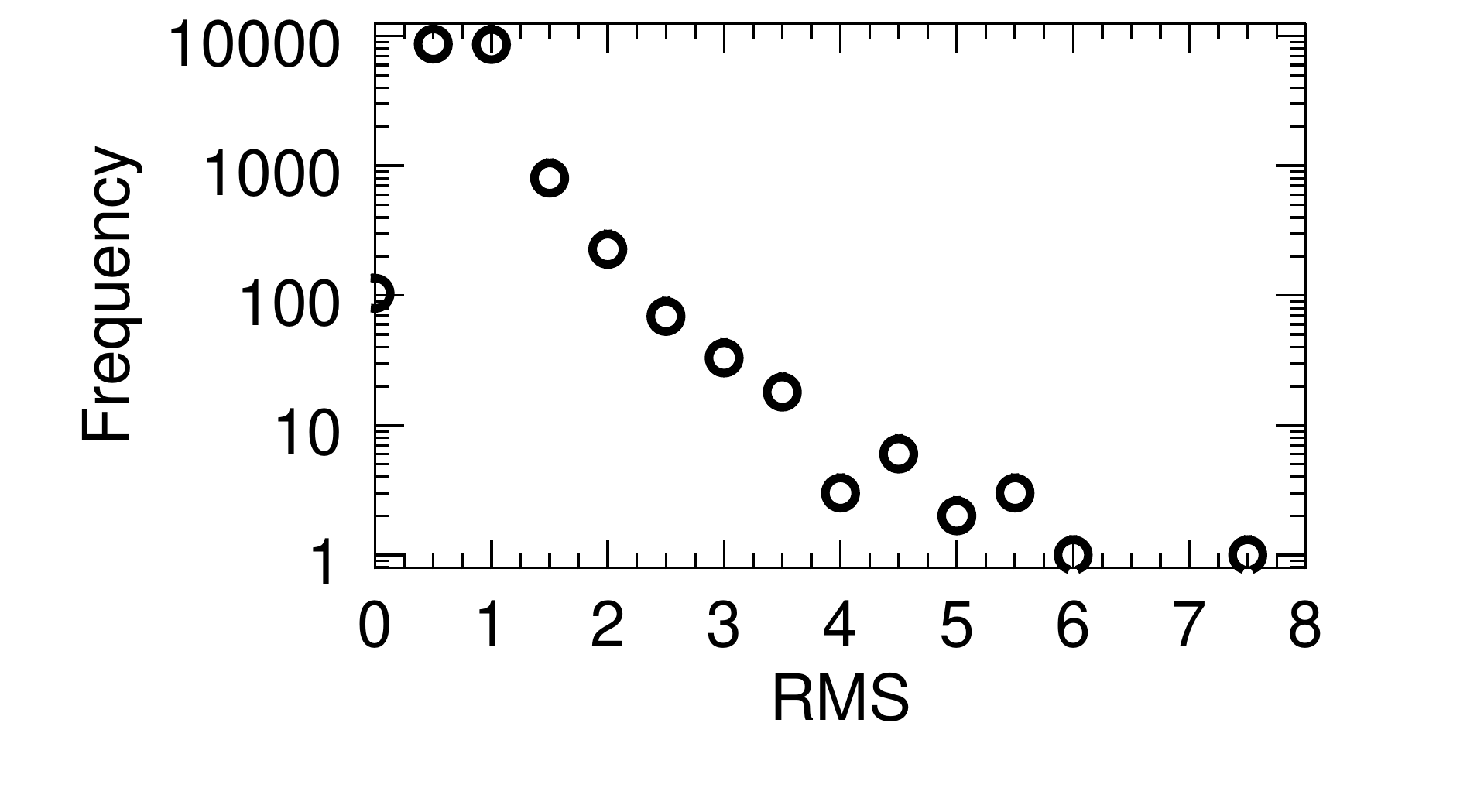}
\caption{Distribution of the root mean square for Case 3 loops.}
\label{fig:Case3b}
\end{subfigure}
\begin{subfigure}{\textwidth}
\hspace{3cm} \includegraphics[width=0.5\textheight, trim = 20 12 0 0, clip]{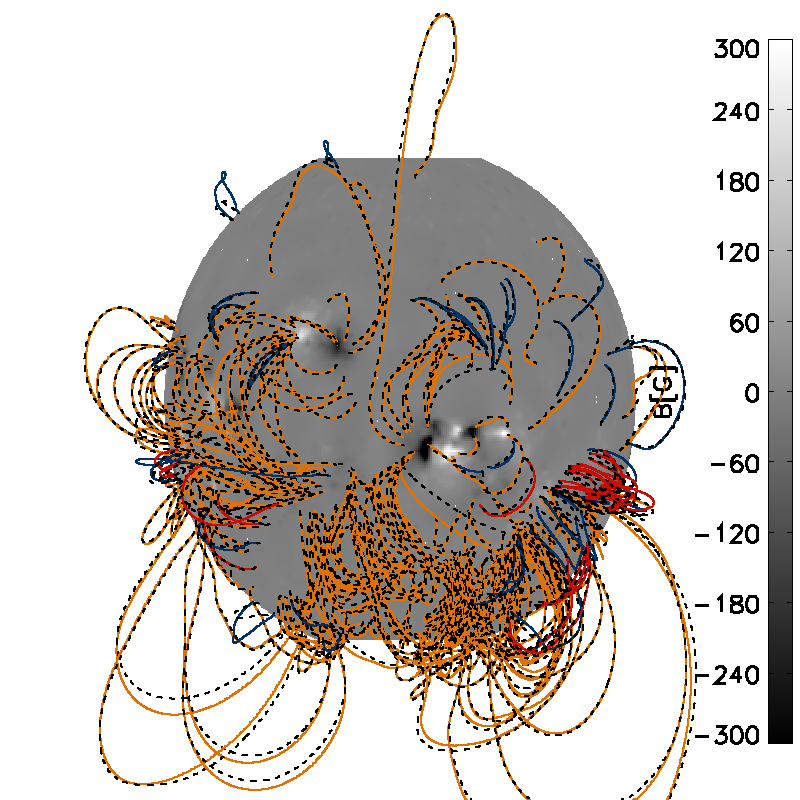}
\caption{Randomly selected 3D loops overplotted on the HMI phototspheric radial magnetic field. The black dashed lines shows the original 3D loops. The CNN reconstructed 3D loops with an RMS (Eq. \ref{RMS}) between 0.5 an 1.5 are displayed with solid orange lines. In blue solid line are shown the 3D loops with an RMS between 0. and 0.5 and between 1.5 and 2. In red solid lines are presented the 3D loops with an RMS larger than 2.}
\label{fig:Case3c}
\end{subfigure}

\caption{CNN results for Data type 2, Case 3.}
\label{fig:image2}
\end{figure*}

For the NLFFF extrapolation we used the computational domain of 180$\times$280$\times$720 grid points extending over r = [1, 2.5] R$_\odot$, $\theta$ = [-70$^\circ$, 70$^\circ$] in latitude and $\phi$ = [0$^\circ$, 360$^\circ$] in longitude. From the solution of the extrapolation, we traced loops higher than r = 1.05 Rs. For the training, we used a total of 18555 loops. We applied to the loops reversible transformations that are useful for the CNN method and which are: coordinate transformations, rotation to ensure z is positive, transformations that x and y components are in the positive domain. We also filtered the loops based on their inflection points to exclude the over wiggling loops which in general are not visible in the EUV images. Compared with the other cases, the Case 3 loops have a wider range in the values of length, distance between footpoints and heights so the CNN method did not perform as well as in the previous cases. In Fig. \ref{fig:Case3a} we show the evolution of the training and validation loss function. For this case, the CNN model converged after 400 epochs. Fig. \ref{fig:Case3b} displays the distribution of the RMS ratios of the loops. From all of the loops, 94.5\% is in the RMS range of 0.5 to 1.5. The performance is not as good as in the previous cases. In Fig. \ref{fig:Case3c} we show the radial magnetic field of the synoptic map at the photospheric level with the disk center corresponding to 180$^\circ$ longitude.

\section{Discussions and conclusions}

Stereoscopy is a suitable method for deriving the 3D geometry of solar phenomena. The study, evolution, and kinematics of these phenomena are essential for understanding their physical processes and the space weather forecasting models.  
This analysis is constrained by the data availability of at least two view directions of the same object, reducing the reconstruction possibilities. On the other hand, solar observations from one view direction are more abundant than two simultaneous views. When only one vantage point is available, the CNN method can be used and applied to coronal loops and other curve-like structures. 

In conclusion, the new 3D loop reconstruction method will allow the user to employ only one image from a single vantage point. One would select as many loops as possible from the EUV image by visual inspection or automatic identification. The 2D curve-like structure parameters mentioned in the Section \ref{CNN_description} can be used as input to the CNN models already trained (see Case 0 ...Case 3 from Section \ref{sec:results}). The output of the used CNN model will be the third component of the 2D curve-like structures. In the end, one will have all the three components of the curve-like structures. 

The neural network model presented in this paper can only perform the 3D geometrical reconstruction of loop types used for the training. It does not have any direct contribution to the magnetic field modeling. Indirectly, the 3D coordinates of the CNN reconstructed loops can be, for example, used as constraint for the magnetic field modeling as shown by \cite{2017ApJ...837...10C}.



\section{Acknowledgements}
Data are courtesy of NASA/SDO and the HMI science teams. IC acknowledge DFG-grant WI 3211/5-1. The HMI data are provided courtesy of NASA/SDO and the HMI science team. I.C. would like to thank M. Madjarska for helpful discussions.
R.G acknowledge financial support by the Portuguese Government throughthe Foundation for Science and Technology - FCT FEDER  -  European  Regional Development Fund through COMPETE 2020 - Operational Programme Competitiveness and Internationalization.

\bibliography{CLRNN.bib}
\bibliographystyle{aasjournal}



\end{document}